\documentstyle[art10]{article}

\author{U.G\"unther\thanks{e-mail: guenther@pool.hrz.htw-zittau.de},
A.Zhuk\thanks{e-mail: zhuk@paco.odessa.ua}\\
Department of Physics, University of Odessa, \\
2 Petra Velikogo St., Odessa 270100, Ukraine}
\title{Gravitational excitons from extra dimensions
}
\def\stackunder#1#2{\mathrel{\mathop{#2}\limits_{#1}}}
\newcommand{\be}[1]{\begin{equation}\label{#1}}
\newcommand{\ee}{\end{equation}}
\newcommand{\ba}[1]{\begin{eqnarray}\label{#1}}
\newcommand{\ea}{\end{eqnarray}}

\font\msbm=msbm10
\def\RR{\hbox{\msbm R}}
\def\NN{\hbox{\msbm N}}

\def\QQ{\hbox{\msbm Q}}

\begin{document}

\maketitle
\renewcommand{\baselinestretch}{1.4} 
$\ \ \ $ \vspace{1cm}\\

\abstract{
Inhomogeneous multidimensional cosmological models with a higher
dimensional space-time manifold $M = M_0\times\prod\nolimits_{i=1}
^nM_i$  $( n \ge 1 )$  are investigated under dimensional reduction
to $D_0$ - dimensional effective models. In the Einstein conformal
frame, small excitations of the scale factors of the internal
spaces near minima of an effective potential have a form of massive
scalar fields in the external space-time. Parameters of models
which ensure  minima  of the effective potentials are obtained for
particular cases and masses of gravitational excitons are estimated.
}

\bigskip

\hspace*{0.950cm} PACS number(s): 04.50.+h, 98.80.Hw 
%

\section{Introduction}

\setcounter{equation}{0}

The large scale dynamics of the observable part of our present time universe
is well described by the Friedmann model with 4-dimensional Friedmann -
Robertson - Walker (FRW) metric. However, it is possible that space-time at
short (Planck) distances might have a dimensionality of more than four and
possess a rather complex topology \cite{1}. String theory \cite{1a} and its
recent generalizations --- p-brane, M- and F-theory \cite{1b,1c} widely use
this concept and give it a new foundation. The most consistent formulations
of these theories are possible in space-times with critical dimensions $%
D_c>4 $, for example, in string theory there are $D_c=26$ or $10$ for the
bosonic and supersymmetric version, respectively. Usually it is supposed
that a $D$-dimensional manifold $M$ undergoes a ''spontaneous
compactification'' \cite{c1}-\cite{c4}: $M\to M^4\times B^{D-4}$ , where $M^4
$ is the 4-dimensional external space-time and $B^{D-4}$ is a compact
internal space. So it is natural to consider cosmological consequences of
such compactifications. With this in mind we shall investigate
multidimensional cosmological models (MCM) with topology 
\begin{equation}
\label{1.1}M=M_0\times M_1\times \dots \times M_n\ , 
\end{equation}
where $M_0$ denotes the $D_0$ - dimensional (usually $D_0=4$) external
space-time and $M_i$ \quad $(i=1,\dots ,n)$ are $D_i$ - dimensional internal
spaces. To make the internal dimensions unobservable at present time these
internal spaces have to be compact and reduced to scales near Planck length $%
L_{Pl}\sim 10^{-33}cm$ , i.e. scale factors $a_i$ of the internal spaces
should be of order of $L_{Pl}$. In this case we cannot move in
extra-dimensions and our space-time is apparently 4-dimensional. There is no
problem to construct compact spaces with a positive curvature \cite{p1,p2}.
(For example, every Einstein manifold with constant positive curvature is
necessarily compact \cite{p3}.) However, Ricci-flat spaces and negative
curvature spaces can be compact also. This can be achieved by appropriate
periodicity conditions for the coordinates \cite{2}-\cite{5a} or,
equivalently, through the action of discrete groups $\Gamma $ of isometries
related to face pairings and to the manifold's topology. For example,
3-dimensional spaces of constant negative curvature are isometric to the
open, simply connected, infinite hyperbolic (Lobachevsky) space $H^3$ \cite
{p1,p2}. But there exist also an infinite number of compact, multiply
connected, hyperbolic quotient manifolds $H^3/\Gamma $ , which can be used
for the construction of FRW metrics with negative curvature \cite{2,4}.
These manifolds are built from a fundamental polyhedron (FP) in $H^3$ with
faces pairwise identified. The FP determines a tessellation of $H^3$ into
cells which are replicas of the FP, through the action of the discrete group 
$\Gamma $ of isometries \cite{4}. The simplest example of Ricci-flat compact
spaces is given by $D$ - dimensional tori $T^D=\RR ^D/\Gamma $ . Thus,
internal spaces may have nontrivial global topology, being compact (i.e.
closed and bounded) for any sign of spatial curvature.

In the cosmological context, internal spaces can be called compactified,
when they are obtained by a compactification \cite{k0} or factorization
(''wrapping'') in the usual mathematical understanding (e.g. by replacements
of the type $\RR ^D\rightarrow S^D$ , $\RR ^D\rightarrow \RR ^D/\Gamma $ or $%
H^D\rightarrow H^D/\Gamma $) with additional contraction of the sizes to
Planck scale. The physical constants that appear in the effective
4-dimensional theory after dimensional reduction of an originally
higher-dimensional model are the result of integration over the extra
dimensions. If the volumes of the internal spaces would change, so would the
observed constants. Because of limitation on the variability of these
constants \cite{k1,k2} the internal spaces are static or at least slowly
variable since the time of primordial nucleosynthesis and as we mentioned
above their sizes are of the order of the Planck length. Obviously, such
compactifications have to be stable against small fluctuations of the sizes
(the scale factors $a_i$) of the internal spaces. This means that the
effective potential of the model obtained under dimensional reduction to a
4-dimensional effective theory should have minima at $a_i\sim L_{Pl}$ $%
(i=1,\dots ,n)$. Because of its crucial role the problem of stable
compactification of extra dimensions was intensively studied in a large
number of papers \cite{6}-\cite{22}. As result certain conditions were
obtained which ensure the stability of these compactifications. However,
position of a system at a minimum of an effective potential means not
necessarily that extra-dimensions are unobservable. As we shall show below,
small excitations of a system near a minimum can be observed as massive
scalar fields in the external space-time. In solid state physics,
excitations of electron subsystems in crystals are called excitons. In our
case the internal spaces are an analog of the electronic subsystem and their
excitations can be called gravitational excitons. If masses of these
excitations are much less than Planck mass $M_{Pl}\sim 10^{-5}g$ , they
should be observable confirming the existence of extra-dimensions. In the
opposite case of very heavy excitons with masses $m\sim M_{Pl}$ it is
impossible to excite them at present time and extra-dimensions are
unobservable by this way.

The paper is organized as follows. In Sec.II we describe our model and
obtain an effective theory in Brans-Dicke and Einstein conformal frames. In
Sec.III it is shown that small excitations of the scale factors of the
internal spaces near minima of an effective potential in the Einstein frame
have a form of massive scalar fields in the external space-time. The masses
of such scalar fields are evaluated for particular classes of effective
potentials with minima in the case of one-internal-space models (Sec.IV) and
two-internal-space models (Sec.V). In Sec.VI we show that conditions for the
existence of stable configurations may be quite different for these two
types of models.

%

\section{The model}

\setcounter{equation}{0}

We consider a cosmological model with metric 
\begin{equation}
\label{2.1}g=g^{(0)}+\sum_{i=1}^ne^{2\beta ^i(x)}g^{(i)}\ , 
\end{equation}
which is defined on manifold (\ref{1.1}) where $x$ are some coordinates of
the $D_0$ - dimensional manifold $M_0$ and 
\begin{equation}
\label{2.2}g^{(0)}=g_{\mu \nu }^{(0)}(x)dx^\mu \otimes dx^\nu \ . 
\end{equation}
Let manifolds $M_i$ be $D_i$ - dimensional Einstein spaces with metric $%
g^{(i)}$ , i.e. 
\begin{equation}
\label{2.3}R_{mn}\left[ g^{(i)}\right] =\lambda ^ig_{mn}^{(i)}\ ,\qquad
m,n=1,\ldots ,D_i 
\end{equation}
and 
\begin{equation}
\label{2.4}R\left[ g^{(i)}\right] =\lambda ^iD_i\equiv R_i\ . 
\end{equation}
In the case of constant curvature spaces parameters $\lambda ^i$ are
normalized as $\lambda ^i=k_i(D_i-1)$ with $k_i=\pm 1,0$. We note that each
of the spaces $M_i$ can be split into a product of Einstein spaces: $%
M_i\rightarrow \prod_{k=1}^{n_i}M_i^k$ \cite{23}. Here $M_i^k$ are Einstein
spaces of dimensions $D_i^k$ with metric $g_{(k)}^{(i)}$: $R_{mn}\left[
g_{(k)}^{(i)}\right] =\lambda _k^ig_{(k)mn}^{(i)}\quad (m,n=1,\ldots ,D_i^k)$
and $R\left[ g_{(k)}^{(i)}\right] =\lambda _k^iD_i^k$. Such a splitting
procedure is well defined provided $M_i^k$ are not Ricci - flat \cite{23,24}%
. If $M_i$ is a split space, then for curvature and dimension we have
respectively \cite{23}: $R\left[ g^{(i)}\right] =\sum_{k=1}^{n_i}R\left[
g_{(k)}^{(i)}\right] $ and $D_i=\sum_{k=1}^{n_i}D_i^k$. Later on we shall
not specify the structure of the spaces $M_i$. We require only $M_i$ to be
compact spaces with arbitrary sign of curvature.

With total dimension $D=\sum_{i=0}^nD_i$, $\kappa ^2$ a $D$ - dimensional
gravitational constant, $\Lambda $ - a $D$ - dimensional cosmological
constant and $S_{YGH}$ the standard York - Gibbons - Hawking boundary term 
\cite{24a,24b}, we consider an action of the form 
\begin{equation}
\label{2.5}S=\frac 1{2\kappa ^2}\int\limits_Md^Dx\sqrt{|g|}\left\{
R[g]-2\Lambda \right\} +S_{add}+S_{YGH}\ . 
\end{equation}
The additional potential term 
\begin{equation}
\label{2.6}S_{add}=-\int\limits_Md^Dx\sqrt{|g|}\rho (x) 
\end{equation}
is not specified and left in its general form, taking into account the
Casimir effect \cite{6}, the Freund - Rubin monopole ansatz \cite{c2}, a
perfect fluid \cite{26,27} or other hypothetical potentials \cite{20,22}. In
all these cases $\rho $ depends on the external coordinates through the
scale factors $a_i(x)=e^{\beta ^i(x)}\ (i=1,\ldots ,n)$ of the internal
spaces. We did not include into the action (\ref{2.5}) a minimally coupled
scalar field with potential $U(\psi )$ , because in this case there exist no
solutions with static internal spaces for scalar fields $\psi $ depending on
the external coordinates \cite{20}.

After dimensional reduction the action reads 
$$
S=\frac 1{2\kappa _0^2}\int\limits_{M_0}d^{D_0}x\sqrt{|g^{(0)}|}%
\prod_{i=1}^ne^{D_i\beta ^i}\left\{ R\left[ g^{(0)}\right] -G_{ij}g^{(0)\mu
\nu }\partial _\mu \beta ^i\,\partial _\nu \beta ^j+\right.  
$$
\begin{equation}
\label{2.7}+\sum_{i=1}^n\left. R\left[ g^{(i)}\right] e^{-2\beta
^i}-2\Lambda -2\kappa ^2\rho \right\} \ ,
\end{equation}
where $\kappa _0^2=\kappa ^2/\mu $ is the $D_0$ - dimensional gravitational
constant, $\mu =\prod_{i=1}^n\mu _i=\prod_{i=1}^n\int\limits_{M_i}d^{D_i}y
\sqrt{|g^{(i)}|}$ and $G_{ij}=D_i\delta _{ij}-D_iD_j\ (i,j=1,\ldots ,n)$ is
the midisuperspace metric \cite{28,29}. Here the scale factors $\beta ^i$ of
the internal spaces play the role of scalar fields. Comparing this action
with the tree-level effective action for a bosonic string it can be easily
seen that the volume of the internal spaces $e^{-2\Phi }\equiv
\prod_{i=1}^ne^{D_i\beta ^i}$ plays the role of the dilaton field \cite
{23,29,30}. We note that sometimes all scalar fields associated with $\beta
^i$ are called dilatons. Action (\ref{2.7}) is written in the Brans - Dicke
frame. Conformal transformation to the Einstein frame 
\begin{equation}
\label{2.8}\hat g_{\mu \nu }^{(0)}=e^{-\frac{4\Phi }{D_0-2}}g_{\mu \nu
}^{(0)}={\left( \prod_{i=1}^ne^{D_i\beta ^i}\right) }^{\frac 2{D_0-2}}g_{\mu
\nu }^{(0)}
\end{equation}
yields 
\begin{equation}
\label{2.9}S=\frac 1{2\kappa _0^2}\int\limits_{M_0}d^{D_0}x\sqrt{|\hat
g^{(0)}|}\left\{ \hat R\left[ \hat g^{(0)}\right] -\bar G_{ij}\hat g^{(0)\mu
\nu }\partial _\mu \beta ^i\,\partial _\nu \beta ^j-2U_{eff}\right\} \ .
\end{equation}
The tensor components of the midisuperspace metric (target space metric on $%
\RR _T^n$ ) $\bar G_{ij}\ (i,j=1,\ldots ,n)$ , its inverse metric $\bar
G^{ij}$ and the effective potential are respectively 
\begin{equation}
\label{2.10}\bar G_{ij}=D_i\delta _{ij}+\frac 1{D_0-2}D_iD_j\ ,
\end{equation}
\begin{equation}
\label{2.11}\bar G^{ij}=\frac{\delta ^{ij}}{D_i}+\frac 1{2-D}
\end{equation}
and 
\begin{equation}
\label{2.12}U_{eff}={\left( \prod_{i=1}^ne^{D_i\beta ^i}\right) }^{-\frac
2{D_0-2}}\left[ -\frac 12\sum_{i=1}^nR_ie^{-2\beta ^i}+\Lambda +\kappa
^2\rho \right] \ .
\end{equation}
We remind that $\rho $ depends on the scale factors of the internal spaces: $%
\rho =\rho \left( \beta ^1,\ldots ,\beta ^n\right) $. Thus, we are led to
the action of a self-gravitating $\sigma -$model with flat target space $(%
\RR _T^n,\bar G)$ (\ref{2.10}) and self-interaction described by the
potential (\ref{2.12}).

Let us first consider the case of one internal space: $n=1$. Redefining the
dilaton field as 
\begin{equation}
\label{2.13}\varphi \equiv \pm \sqrt{\frac{D_1(D-2)}{D_0-2}}\beta ^1 
\end{equation}
we get for action and effective potential respectively 
\begin{equation}
\label{2.14}S=\frac 1{2\kappa _0^2}\int d^{D_0}x\sqrt{|\hat g^{(0)}|}\left\{
\hat R\left[ \hat g^{(0)}\right] -\hat g^{(0)\mu \nu }\partial _\mu \varphi
\,\partial _\nu \varphi -2U_{eff}\right\} 
\end{equation}
and 
\begin{equation}
\label{2.15}U_{eff}=e^{2\varphi {\left[ \frac{D_1}{(D-2)(D_0-2)}\right] }%
^{1/2}}\left[ -\frac 12R_1e^{2\varphi {\left[ \frac{D_0-2}{D_1(D-2)}\right] }%
^{1/2}}+\Lambda +\kappa ^2\rho (\varphi )\right] \ , 
\end{equation}
where in the latter expression we use for definiteness sign minus.

Coming back to the general case $n>1$ we bring midisuperspace metric (target
space metric) (\ref{2.10}) by a regular coordinate transformation 
\begin{equation}
\label{2.16}\varphi =Q\beta \quad ,\quad \beta =Q^{-1}\varphi \ 
\end{equation}
to a pure Euclidean form 
\begin{equation}
\label{2.17} 
\begin{array}{c}
\bar G_{ij}d\beta ^i\otimes d\beta ^j=\sigma _{ij}d\varphi ^i\otimes
d\varphi ^j=\sum_{i=1}^nd\varphi ^i\otimes d\varphi ^i\ , \\  
\\ 
\bar G=Q^{\prime }Q\quad ,\quad \sigma ={\rm diag\ }(+1+1,\ldots ,+1)\ . 
\end{array}
\end{equation}
(The prime denotes the transposition.) An appropriate transformation $Q:\
\beta ^i\mapsto \varphi ^j=Q_i^j\beta ^i$ is given e.g. by \cite{28} 
\begin{eqnarray}\label{2.18}
\varphi ^1 & = & - A\sum_{i=1}^nD_i\beta ^i\ , \nonumber\\
\varphi ^i & = & {\left[\left.D_{i-1}\right/\Sigma _{i-1}\Sigma
_i\right]}^{1/2}\sum_{j=i}^{n}D_j\left(\beta ^j - \beta
^{i-1}\right)\ ,\quad i = 2,\ldots ,n\ ,
\end{eqnarray} where $\Sigma _i=\sum_{j=i}^nD_j$ , 
\begin{equation}
\label{2.20}A=\pm {\left[ \frac 1{D^{\prime }}\frac{D-2}{D_0-2}\right] }%
^{1/2}\ , 
\end{equation}
and $D^{\prime }=\sum_{i=1}^nD_i$. So we can write action (\ref{2.9}) as 
\begin{equation}
\label{2.21}S=\frac 1{2\kappa _0^2}\int\limits_{M_0}d^{D_0}x\sqrt{|\hat
g^{(0)}|}\left\{ \hat R\left[ \hat g^{(0)}\right] -\sigma _{ik}\hat
g^{(0)\mu \nu }\partial _\mu \varphi ^i\,\partial _\nu \varphi
^k-2U_{eff}\right\} 
\end{equation}
with effective potential 
\begin{equation}
\label{2.22}U_{eff}=e^{\frac 2{A(D_0-2)}\varphi ^1}\left( -\frac
12\sum_{i=1}^nR_ie^{-2{(Q^{-1})^i}_k\varphi ^k}+\Lambda +\kappa ^2\rho
\right) \ . 
\end{equation}

%

\section{Gravitational excitons\label{mark1}}

\setcounter{equation}{0}

Let us suppose that the effective potential (\ref{2.12}) has minima at
points $\vec \beta _c=\left( \beta _c^1,\ldots ,\beta _c^n\right) $ 
\begin{equation}
\label{3.1}\left. \frac{\partial U_{eff}}{\partial \beta ^i}\right| _{\vec
\beta _c}=0\ ,\quad c=1,\ldots ,m\ 
\end{equation}
and that its Hessian 
\begin{equation}
\label{3.2}a_{(c)ik}:=\left. \frac{\partial ^2U_{eff}}{\partial \beta
^i\,\partial \beta ^k}\right| _{\vec \beta _c}
\end{equation}
does not identically vanish at these points. For small fluctuations $\eta
^i:=\beta ^i-\beta _c^i$ we have then up to second order in the Taylor
expansion 
\begin{equation}
\label{3.3}U_{eff}=U_{eff}\left( \vec \beta _c\right) +\frac
12\sum_{i,k=1}^na_{(c)ik}\eta ^i\eta ^k\ .
\end{equation}
As sufficient condition for the existence of minima at $\vec \beta _c$ we
choose in this paper the strong condition consisting in the positivity of
the quadratic form 
\begin{equation}
\label{3.4}{\eta }^{\prime }A_c\eta \equiv \sum_{i,k=1}^na_{(c)ik}\eta
^i\eta ^k>0\ ,\quad \forall \eta ^1,\ldots ,\eta ^n\ ,
\end{equation}
with exception of the point $\eta ^1=\eta ^2=\ldots =\eta ^n=0$. It is clear
that for higher order expansions of the effective potential inequality (\ref
{3.4}) can be weaken to a nonnegativity condition ${\eta }^{\prime }A_c\eta
\geq 0$ with additional requirements on the multilinearforms occurring in
this case. We note that according to the Sylvester criterion positivity of
quadratic forms is assured by the positivity of the principal minors of the
corresponding matrix, in our case of the matrix $A_c:$%
\begin{equation}
\label{3.4a}
\begin{array}{l}
a_{(c)11}>0\ ,\quad \left| 
\begin{array}{cc}
a_{(c)11} & a_{(c)12} \\ 
a_{(c)21} & a_{(c)22}
\end{array}
\right| >0\ ,\quad \ldots  \\  
\\ 
\ldots \ ,\quad \left| 
\begin{array}{ccc}
a_{(c)11} & \cdots  & a_{(c)1n} \\ 
a_{(c)21} & \cdots  & a_{(c)2n} \\ 
\cdots  & \cdots  & \cdots  \\ 
a_{(c)n1} & \cdots  & a_{(c)nn}
\end{array}
\right| =\det A_c>0\ .
\end{array}
\end{equation}

Equation (\ref{3.1}) and Hessian (\ref{3.2}) are affected by midisuperspace
coordinate transformation (\ref{2.16}) as follows: 
\begin{equation}
\label{3.5}\left. \frac{\partial U_{eff}}{\partial \varphi ^i}\right| _{\vec
\varphi _c}=\left. \frac{\partial U_{eff}}{\partial \beta ^k}\right| _{\vec
\beta _c}{(Q^{-1})^k}_i=0\ ,\quad \varphi _c=Q\beta _c\ , 
\end{equation}
\begin{equation}
\label{3.6}a_{(c)ik}=\left. \frac{\partial ^2U_{eff}}{\partial \beta
^i\,\partial \beta ^k}\right| _{\vec \beta _c}=\frac{\partial \varphi ^j}{%
\partial \beta ^i}\left. \frac{\partial ^2U_{eff}}{\partial \varphi
^j\,\partial \varphi ^l}\right| _{\vec \varphi _c}\frac{\partial \varphi ^l}{%
\partial \beta ^k}\equiv Q_i^j\bar a_{(c)jl}Q_k^l\ . 
\end{equation}
This means that matrices $A_c$ and $\bar A_c$ are congruent matrices \cite
{31} $A_c=Q^{\prime }\bar A_cQ$ , and hence their rank and signature
coincide.

Taking into account that transformation (\ref{2.16}) holds also for small
fluctuations near the minima 
\begin{equation}
\label{3.7}\xi =Q\eta \ ,\quad \eta =Q^{-1}\xi \ ,\quad \xi ^i:=\varphi
^i-\varphi _c^i 
\end{equation}
we conclude that the quadratic form (\ref{3.4}) is invariant under this
transformation 
\begin{equation}
\label{3.8}{\eta }^{\prime }A_c\eta =(Q^{-1}\xi )^{\prime }Q^{\prime }\bar
A_cQ(Q^{-1}\xi )=\xi ^{\prime }\bar A_c\xi \ . 
\end{equation}
Together with the coinciding rank and signature of the congruent matrices $%
\bar A_c$ and $A_c$ this implies that the positivity of (\ref{3.4}) remains
preserved and minima of $U_{eff}$ in $\beta -$representation correspond to
minima of $U_{eff}$ in $\varphi -$representation.

To get masses of excitations we need to diagonalize the matrices $\bar A_c$
, keeping at the same time the kinetic term $\hat g^{(0)\mu \nu
}\sum_{i=1}^n\varphi _{,\mu }^i\varphi _{,\nu }^i$ in its diagonal form. One
immediately checks that appropriate $SO(n)-$rotations $S_c:\ S_c^{\prime
}=S_c^{-1}$ fulfill these requirements 
\begin{equation}
\label{3.9}\bar A_c=S_c^{\prime }M_c^2S_c\ ,\quad M_c^2={\rm diag\ }%
(m_{(c)1}^2,m_{(c)2}^2,\ldots ,m_{(c)n}^2) 
\end{equation}
and 
\begin{equation}
\label{3.10}\hat g^{(0)\mu \nu }\sum_{i=1}^n\varphi _{,\mu }^i\varphi _{,\nu
}^i=\hat g^{(0)\mu \nu }\sum_{i=1}^n\phi _{,\mu }^i\phi _{,\nu }^i\ , 
\end{equation}
where $\phi =S_c\varphi $. Introducing corresponding transformed fluctuation
fields $\psi =S_c\xi $ we also verify that 
\begin{equation}
\label{3.11}\eta ^{\prime }A_c\eta =\xi ^{\prime }\bar A_c\xi =\psi ^{\prime
}M_c^2\psi \ . 
\end{equation}
It is clear from the Sylvester criterion that all diagonal elements of the
matrix $M_c^2$ should be positive. From relations (\ref{2.17}), (\ref{3.6}),
(\ref{3.9}) follows that they are eigenvalues of matrix $\bar A_c$ as well
as matrix $\bar G^{-1}A_c$ .

So explicit calculations of the matrices $S_c$ and $M_c^2$ go along standard
lines \cite{31} and give e.g. in the case of two internal spaces $(n=2)$ 
\begin{equation}
\label{3.15}{S_c}=\left( 
\begin{array}{cc}
\cos {\alpha _c} & -\sin 
{\alpha _c} \\ \sin {\alpha _c} & \cos {\alpha _c} 
\end{array}
\right) 
\end{equation}
with the angle of rotation 
\begin{equation}
\label{3.16}\tan {2\alpha _c}=\frac{2\bar a_{(c)12}}{\bar a_{(c)22}-\bar
a_{(c)11}}\ 
\end{equation}
and 
\begin{equation}
\label{3.17}m_{(c)1,2}^2=\frac 12\left[ Tr(B_c)\pm \sqrt{Tr^2(B_c)-4\det
(B_c)}\right] \ , 
\end{equation}
where 
\begin{equation}
\label{3.18}B_c=\bar A_c\quad {\rm or}\quad B_c=\bar G^{-1}A_c\ . 
\end{equation}
It can be easily seen that $m_{(c)1}^2\ ,\ m_{(c)2}^2$ are positive because $%
\bar a_{(c)11},\bar a_{(c)22}>0$ and $\bar a_{(c)11}\bar a_{(c)22}>\bar
a_{(c)12}^2$\ .

So, the action functional (\ref{2.21}) is equivalent to a family of action
functionals for small fluctuations of the scale factors of internal spaces
in the vicinity of the minima of the effective potential 
\begin{eqnarray}\label{3.19}
S & = & \frac{1}{2\kappa _0^2}\int \limits_{M_0}d^{D_0}x \sqrt
{|\hat g^{(0)}|}\left\{\hat R\left[\hat g^{(0)}\right] - 2\Lambda
_{(c)eff}\right\} + \nonumber\\
\ & + & \sum_{i=1}^{n}\frac{1}{2}\int \limits_{M_0}d^{D_0}x \sqrt
{|\hat g^{(0)}|}\left\{-\hat g^{(0)\mu \nu}\psi ^i_{,\mu}\psi
^i_{,\nu} -
m_{(c)i}^2\psi ^i\psi ^i\right\}\ , \quad c=1,\ldots ,m\ ,
\end{eqnarray}where $\Lambda _{(c)eff}:=U_{eff}\left( \vec \phi _c\right) $
and the factor $\sqrt{\mu /\kappa ^2}$ has been included into $\psi $ for
convenience: $\sqrt{\mu /\kappa ^2}\psi \rightarrow \psi $.

Thus, conformal excitations of the metric of the internal spaces behave as
massive scalar fields developing on the background of the external space -
time. By analogy with excitons in solid state physics where they are
excitations of the electronic subsystem of a crystal, the excitations of the
internal spaces may be called gravitational excitons.

In the conclusion of this section we want to make a few remarks concerning
the form of the effective potential. From the physical viewpoint it is clear
that the effective potential should provide following conditions:%
\begin{eqnarray}\label{3.19a}
(i)\, \qquad a_{(c)i} & = & e^{\beta_c^i}\,  \mbox{ \small
$^{>}_{\sim}$ }\,L_{Pl}\ ,
\nonumber\\
(ii)\ \ \, \quad m_{(c)i} & \leq & M_{Pl}\ ,\nonumber\\
(iii) \,  \quad  \Lambda _{(c)eff} & \rightarrow & 0\ .
\end{eqnarray}
The first condition expresses the fact that the internal spaces should be
unobservable at the present time and stable against quantum gravitational
fluctuations.This condition ensures the applicability of the classical
gravitational equations near positions of minima of the effective potential.
The second condition means that the curvature of the effective potential
should be less than Planckian one. Of course, gravitational excitons can be
excited at the present time if $m_i\ll M_{Pl}$. The third condition reflects
the fact that the cosmological constant at the present time is very small: $%
\Lambda \leq 10^{-54}{\mbox{cm}}^{-2}\approx 10^{-120}\Lambda _{Pl}$ where $%
\Lambda _{Pl}=L_{Pl}^{-2}$. Thus, for simplicity, we can demand $\Lambda
_{eff}=U_{eff}(\vec \beta _c)=0$. (We used the abbreviation $\Lambda
_{eff}:=\Lambda _{(c)eff}$ .) Strictly speaking, in the multi-minimum case $%
(c>1)$ we can demand $a_{(c)i}\sim L_{Pl}$ and $\Lambda _{(c)eff}=0$ only
for one of the minima, namely the minimum which corresponds to the state of
the present universe. For all other minima it may be $a_{(c)i}\gg L_{Pl}$
and $|\Lambda _{(c)eff}|\gg 0$.

It can be easily seen that the conditions $\Lambda _{eff}=0$ and $\rho
\equiv 0$ are incompatible. In fact, the necessary extremum - condition for
the potential (\ref{2.22}) reads 
\begin{eqnarray}\label{3.20}
\tilde B^{-1}\cdot \frac{\partial U_{eff}}{\partial \varphi ^1} & = &
\sum_{j=1}^{n}r_j{(Q^{-1})^j}_1 + \frac{\partial \rho}{\partial
\varphi ^1}
+ q_1\tilde B^{-1}U_{eff} = 0\ , \nonumber\\
\tilde B^{-1}\cdot \frac{\partial U_{eff}}{\partial \varphi ^i} & = &
\sum_{j=1}^{n}r_j{(Q^{-1})^j}_i + \frac{\partial \rho}{\partial
\varphi ^i}
= 0\ , i=2,\ldots ,n\ ,
\end{eqnarray}
where $r_i:=R_i\exp {\left( -2{(Q^{-1})^i}_k\varphi ^k\right) }\ ,\quad
\tilde B:=\exp {q_1\varphi ^1}$ and $q_1=2/A(D_0-2)$. For $\left.
U_{eff}\right| _{min}=0$ and $\rho \equiv 0$ this system has a nontrivial
solution iff $\det Q=0$. But transformation (\ref{2.16}) is regular. Thus,
there are no solutions for $\left. U_{eff}\right| _{min}=0$ and $\rho \equiv
0$ unless all internal spaces are Ricci - flat. Moreover, as follows from
potential (\ref{2.12}), the conditions $\left. U_{eff}\right| _{min}=0$ and $%
\left. \frac{\partial U_{eff}}{\partial \beta ^i}\right| _{min}=0$ are
compatible iff 
\begin{equation}
\label{3.21}\sum_{i=1}^nR_ie^{-2\beta _c^i}=2\left( \Lambda +\kappa ^2\rho
\left( \vec \beta _c\right) \right) 
\end{equation}
and 
\begin{equation}
\label{3.22}R_ie^{-2\beta _c^i}=-\kappa ^2\left. \frac{\partial \rho }{%
\partial \beta ^i}\right| _{\vec \beta _c}\ ,i=1,\ldots ,n\ . 
\end{equation}
If all internal spaces are Ricci - flat $(R_i\equiv 0,\ i=1,\ldots ,n)$ and $%
\rho \equiv 0$, there are no extrema at all.

With $\left. U_{eff}\right| _{\vec \beta _c}=0$ , $\left. \frac{\partial
U_{eff}}{\partial \beta ^i}\right| _{\vec \beta _c}=0$ and eq. (\ref{3.22})
the Hessian (\ref{3.2}) of the potential (\ref{2.12}) reads 
\begin{equation}
\label{3.23}a_{(c)ik}=\bar B\kappa ^2\left[ 2\delta _{ik}\left. \frac{%
\partial \rho }{\partial \beta ^i}\right| _{\vec \beta _c}+\left. \frac{%
\partial ^2\rho }{\partial \beta ^i\,\partial \beta ^k}\right| _{\vec \beta
_c}\right] \ , 
\end{equation}
where $\bar B:=\exp {\left[ -\frac 2{D_0-2}\sum_{i=1}^nD_i\beta _c^i\right] }
$. The effective potential $U_{eff}$ has minima at $\vec \beta _c$ if
matrices $a_{(c)ik}$ satisfy the Sylvester criterion (\ref{3.4}). Because of 
$\bar B>0$, it is sufficient to check this criterion for the matrix elements 
$h_{ij}=\bar B^{-1}a_{ij}$. For example, in the two - internal - spaces case 
$(n=2)$ there will be minima if 
\begin{equation}
\label{3.24}2\left. \frac{\partial \rho }{\partial \beta ^i}\right| _{\vec
\beta _c}+\left. \frac{\partial ^2\rho }{\partial {\beta ^i}^2}\right|
_{\vec \beta _c}>0\ ,\quad i=1,2\ 
\end{equation}
and 
\begin{equation}
\label{3.25}\prod_{i=1}^2\left( 2\left. \frac{\partial \rho }{\partial \beta
^i}\right| _{\vec \beta _c}+\left. \frac{\partial ^2\rho }{\partial {\beta ^i%
}^2}\right| _{\vec \beta _c}\right) >{\left( \left. \frac{\partial ^2\rho }{%
\partial \beta ^1\,\partial \beta ^2}\right| _{\vec \beta _c}\right) }^2\ . 
\end{equation}
Let us suppose a structure of $\rho $: 
\begin{equation}
\label{3.26}\rho =\sum_{a=1}^NA_a\exp {\left( \sum_{k=1}^n{f^a}_k\beta
^k\right) }\ , 
\end{equation}
where $A_a,{f^a}_k$ are constants. This potential has very general form and
includes, for example, Freund - Rubin ''monopole'' ansatz \cite{c2}, crude
approximations of the Casimir effect due to non-trivial topology of the
space-time \cite{6,22} and multicomponent perfect fluids \cite{26,27}. In
the former case (''monopole'') the potential $\rho $ reads \cite{11} 
\begin{equation}
\label{3.27}\rho =\sum_{i=1}^n\frac{{(f_i)}^2}{a_i^{2D_i}}=\sum_{i=1}^n{(f_i)%
}^2e^{-2D_i\beta ^i}\ , 
\end{equation}
where $f_i$ = const. So, for the matrix ${f^i}_k$ we have ${f^i}%
_k=-2D_i\delta _{ik}\ ,\quad i,k=1,\ldots ,n$. In the case of the
multicomponent perfect fluid the energy density reads \cite{26,27} 
\begin{equation}
\label{3.30}\rho =\sum_{a=1}^m\rho ^{(a)}=\sum_{a=1}^mA_a\exp {\left(
-\sum_{k=1}^n\alpha _k^{(a)}D_k\beta ^k\right) }\ , 
\end{equation}
where $A_a$ are constants. This formula describes the $m$ - component
perfect fluid with the equations of state $P_i^{(a)}=\left( \alpha
_i^{(a)}-1\right) \rho ^a$ in the internal space $M_i\ (i=1,\ldots ,n)$. In
the external space each component corresponds to vacuum: $\alpha
_0^{(a)}=0\,(a=1,\ldots ,m)$. For this example ${f^a}_k=-\alpha _k^{(a)}D_k$.

For the potential (\ref{3.26}) equation (\ref{3.22}) can be rewritten as 
\begin{equation}
\label{3.31}r_k=-\kappa ^2\sum_{a=1}^Nh_a{f^a}_k\ ,\quad k=1,\ldots ,n\ , 
\end{equation}
where $r_k:=R_k\exp {\left( -2\beta _c^k\right) }$ and $h_a:=A_a\exp {\left(
\sum_{k=1}^n{f^a}_k\beta _c^k\right) }$. Now, the minimum - conditions (\ref
{3.24}) and (\ref{3.25}) respectively read 
\begin{equation}
\label{3.32}\sum_{a=1}^Nh_a{{f^a}_k\left( {f^a}_k+2\right) }>0\ ,\quad k=1,2 
\end{equation}
and 
\begin{equation}
\label{3.33}\prod_{k=1}^2\left( \sum_{a=1}^Nh_a{{f^a}_k\left( {f^a}%
_k+2\right) }\right) >\left( \sum_{a=1}^Nh_a{f^a}_1{f^a}_2\right) ^2\ . 
\end{equation}
For example, for the ''monopole'' potential (\ref{3.27}) we obtain the
extremum - condition: 
\begin{equation}
\label{3.34}R_k\exp {\left( 2(D_k-1)\beta _c^k\right) }=2D_k\kappa ^2{(f_k)}%
^2\ ,\quad k=1,\ldots ,n\ . 
\end{equation}
It follows from this expression that there exists an extremum if $%
\mbox{\rm sign}\,R_k>0\ ,\quad k=1,\ldots ,n$ . Conditions (\ref{3.32}), (%
\ref{3.33}) show that this extremum is a minimum (for $D_k>1$).

%

\section{One internal space\label{mark2}}

\setcounter{equation}{0}

Here we consider the case of one internal space or, strictly speaking, the
case where all internal spaces have one common scale factor. In the case
under consideration the action and the effective potential are given by
equations (\ref{2.14}) and (\ref{2.15}) respectively. To get masses of the
gravitational excitons it is necessary to specify the potential $\rho $ .
For this purpose we consider four particular examples:

a) pure geometrical potential: $\rho \equiv 0$.

The necessary condition for the existence of an extremum gives 
\begin{equation}
\label{4.1}\frac{R_1}{D_1}e^{-2\beta _c}=\frac{2\Lambda }{D-2}\ ,
\end{equation}
where $\beta :=\beta ^1$. It follows from this expression that $%
\mbox{\rm sign}\,\Lambda =\mbox{\rm sign}\,R_1$. From the minimum-condition 
\begin{equation}
\label{4.2}a_{11}=\left. \frac{\partial ^2U_{eff}}{\partial \beta ^2}\right|
_{\beta _c}=-\frac{2(D-2)}{D_0-2}R_1{\left( e^{-2\beta _c}\right) }^{\frac{%
D-2}{D_0-2}}>0\ 
\end{equation}
we see that bare cosmological constant and curvature of the internal space
should be negative: $\Lambda ,R_1<0$. The effective cosmological constant is 
\begin{equation}
\label{4.3}\Lambda _{eff}=\frac{D_0-2}{2D_1}R_1{\left( e^{-2\beta _c}\right) 
}^{\frac{D-2}{D_0-2}}
\end{equation}
and negative for $R_1<0$. The mass squared of the exciton reads 
\begin{equation}
\label{4.4}m^2=-\frac{4\Lambda _{eff}}{D_0-2}=\frac{2|R_1|}{D_1}{\left( {%
e^{-2\beta _c}}\right) }^{\frac{D-2}{D_0-2}}\ .
\end{equation}
If we assume, for example, that for a space-time configuration $M_0\times M_1
$ with four-dimensional external space-time $(D_0=4)$ and compact internal
factor-space $M_1=H^{D_1}/\Gamma $ with constant negative curvature $%
R_1=-D_1(D_1-1)$ there exists a minimum of the effective potential at $%
a_c=10^2L_{Pl}$ then we get $m^2=2(D_1-1)10^{-2(D_1+2)}M_{Pl}^2$ and $%
\Lambda _{eff}=-(D_1-1)10^{-2(D_1+2)}\Lambda _{Pl}$ . Thus, according to
observational data with $\left| \Lambda _{eff}\right| \leq 10^{-120}\Lambda
_{Pl}$ there should be at least $D_1=59$ and the corresponding excitons
would be extremely light particles with masses $m\leq 10^{-60}M_{Pl}\sim
10^{-55}$g. If one uses an reduction of the effective cosmological constant
holding $\Lambda =2R_1$ and $R_1$ fixed when $D_1\rightarrow \infty $ (this
can be achieved by a conformal transformation $g^{(1)}\rightarrow
D_1^2g^{(1)}$ with fixed $\kappa _0^2=\kappa ^2/\mu $) one gets $%
a_c\rightarrow L_{Pl}$ and $\Lambda _{eff}\rightarrow 0$ . But at the same
time the exciton mass vanishes $(m\rightarrow 0)$ and the effective
potential degenerates into a step function with infinite height: $%
U_{eff}\rightarrow \infty $ for $a<1$ and $U_{eff}=0$ for $a\geq 1$. Thus,
in the limit $D_1\rightarrow \infty $ there is no minimum at all.

As it was shown in the previous section the effective cosmological constant
is not equal to zero if $\rho \equiv 0$. To satisfy this condition we should
consider the case $\rho \not \equiv 0$.

b) Casimir potential: $\rho =Ce^{-D\beta }$ .

Because of a nontrivial topology of the space - time, vacuum fluctuations of
quantized fields result in a non - zero energy density of the form \cite
{6,9,12,32,33} 
\begin{equation}
\label{4.5}\rho =Ce^{-D\beta }\ ,
\end{equation}
where $C$ is a constant and its value depends strongly on the topology of
the model. For example, for fluctuations of scalar fields the constant $C$
was calculated to take the values: $C=-8.047\cdot 10^{-6}$ if $M_0=\RR %
\times S^3\ ,\ M_1=S^1$ (with $e^{\beta ^0}$ as scale factor of $S^3$ and $%
e^{\beta ^0}\gg e^{\beta ^1}$) \cite{9}; $C=-1.097$ if $M_0=\RR \times \RR %
^2\ ,\ M_1=S^1$ \cite{32} and $C=3.834\cdot 10^{-6}$ if $M_0=\RR \times S^3\
,\ M_1=S^3$ (with $e^{\beta ^0}\gg e^{\beta ^1}$) \cite{9}.

From equations (\ref{3.21}) and (\ref{3.22}) (for $n=1$), i.e. conditions $%
\left. \frac{\partial U_{eff}}{\partial \beta }\right| _{min}=0$ and $%
\Lambda _{eff}=0$ , we immediately derive 
\begin{equation}
\label{4.6}R_1e^{-2\beta _c}=\frac{2D}{D-2}\Lambda 
\end{equation}
and 
\begin{equation}
\label{4.7}R_1e^{(D-2)\beta _c}=\kappa ^2CD\ . 
\end{equation}
An extremum exists if $\mbox{\rm sign}\,R_1=\mbox{\rm sign}\,\Lambda =%
\mbox{\rm sign}\,C$. The expressions (\ref{4.6}) and (\ref{4.7}) provide
fine tuning for the parameters of the model. Similar fine tuning was
obtained by different methods in papers \cite{12} (for one internal space)
and \cite{20} (for $n$ identical internal spaces). The second derivative and
mass squared read respectively 
\begin{eqnarray}\label{4.8}
a_{11} & = & \left.\frac{\partial ^2U_{eff}}{\partial \beta
^2}\right|_{\beta _c} = (D-2)R_1
{\left(e^{-2\beta _c}\right)}^{\frac{D-2}{D_0-2}}\ , \\
m^2 & = & \frac{D_0-2}{D_1}R_1
{\left(e^{-2\beta _c}\right)}^{\frac{D-2}{D_0-2}}\ .
\end{eqnarray}Thus, the internal space should have positive curvature: $%
R_1>0 $ (or for split space $M_1$ the sum of the curvatures of the
constituent spaces $M_1^k$ should be positive).

Let us consider a manifold $M$ with topology $M=\RR
\times S^3\times S^3$ where $e^{\beta ^0}\gg e^{\beta ^1}$. Then \cite{9}, $%
C=3.834\cdot 10^{-6}>0$. As $C,R_1>0$, the effective potential has a minimum
provided $\Lambda >0$. Normalizing $\kappa _0^2$ to unity, we get $\kappa
^2=\mu $ where $\mu =2\left. \pi ^{(d+1)/2}\right/ \Gamma \left( \frac
12(d+1)\right) $ for the $d$ - dimensional sphere. For the model under
consideration we obtain $a_c\approx 1.5\cdot 10^{-1}L_{Pl}$ and $m\approx
2.12\cdot 10^2M_{Pl}$. Hence, the conditions (i) and (ii) are not satisfied
for this topology. For other topologies this problem needs a separate
investigation.

c) ''monopole'' potential: $\rho =f^2e^{-2D_1\beta }$ .

The ''monopole'' ansatz \cite{c2} consists in the proposal that an
antisymmetric tensor field of rank $D_1$ is not equal to zero only for
components corresponding to the internal space $M_1$. The energy density of
this field reads \cite{10,11} 
\begin{eqnarray}\label{4.10}
\rho = f^2e^{-2D_1\beta}\ ,
\end{eqnarray}where $f$ is an arbitrary constant (free parameter of the
model). The equations (\ref{3.21}), (\ref{3.22}) and (\ref{3.34}) yield
following zero - extremum - conditions: 
\begin{equation}
\label{4.11}\Lambda =\frac{D_1-1}{2D_1}R_1e^{-2\beta _c} 
\end{equation}
and 
\begin{equation}
\label{4.12}\frac{R_1}{2D_1\kappa ^2f^2}=e^{-2\beta _c(D_1-1)}\ , 
\end{equation}
which show that $R_1,\Lambda >0$. The exciton mass squared reads 
\begin{eqnarray}\label{4.13}
m^2 = \frac{2(D_0-2)(D_1-1)}{D_1(D-2)}R_1
{\left (e^{-2\beta _c}\right)}^{\frac{D-2}{D_0-2}}\ .
\end{eqnarray} Condition (i) is satisfied if 
\begin{eqnarray}\label{4.14}
f^2  \mbox{ \small $^{>}_{\sim}$ }\left.R_1\right/2\kappa
^2D_1\ .
\end{eqnarray}Let $M_1$ be a 3 - dimensional sphere, then $R_1=6$ and $%
\kappa ^2=2\pi ^2$. To get a minimum of the effective potential for a scale
factor $a_c=10L_{Pl}$ we should take $f^2\approx 5\cdot 10^2$. For this
value of $a_c$ and for $D_0=4$ the mass squared is $m^2= \frac{16}5\cdot
10^{-5}\ll M_{Pl}^2$. Thus, all three conditions (i) - (iii) are satisfied.

d) perfect - fluid potential: $\rho = Ae^{-\alpha D_1\beta}$ .

The one - component perfect - fluid potential reads \cite{26,27} 
\begin{eqnarray}\label{4.15}
\rho = Ae^{-\alpha D_1\beta}\ ,
\end{eqnarray}where $A$ is an arbitrary positive constant. It describes
vacuum in the external space and a perfect fluid with the equation of state $%
P=(\alpha -1)\rho $ in the internal space $M_1$. Physical values of $\alpha $
are restricted to 
\begin{eqnarray}\label{4.16}
0\leq \alpha \leq 2\ .
\end{eqnarray}It is easy to see that the case $\alpha =0$ corresponds to the
vacuum in the space $M_1$ and contributes to the bare cosmological constant $%
\Lambda $. Therefore we shall not consider $\alpha =0$ because in this case
we come back to subsection a). The other limiting case with $\alpha =2$
formally coincides here with the ''monopole'' potential (\ref{4.10}).

For the perfect - fluid potential (\ref{4.15}) a vanishing effective
cosmological constant $\Lambda _{eff}=0$ (eq. (\ref{3.21})) and extremum
condition (\ref{3.22}) yield 
\begin{equation}
\label{4.17}R_1e^{(\alpha D_1-2)\beta _c}=\kappa ^2\alpha D_1A 
\end{equation}
and 
\begin{equation}
\label{4.18}R_1e^{-2\beta _c}=\frac{2\alpha D_1}{\alpha D_1-2}\Lambda \ . 
\end{equation}
For the second derivative of the effective potential in the minimum we
obtain: 
\begin{equation}
\label{4.19}a_{11}=\left. \frac{\partial ^2U_{eff}}{\partial \beta ^2}%
\right| _{\beta _c}=(\alpha D_1-2)R_1{\left( e^{-2\beta _c}\right) }^{\frac{%
D-2}{D_0-2}}\ . 
\end{equation}
Because of $\alpha ,A>0$, equation (\ref{4.17}) shows that the internal
space $M_1$ should have a positive curvature: $R_1>0$. From eq. (\ref{4.19})
we see that there exists a minimum if $\alpha >2/D_1$ . The corresponding
mass squared of the exciton is given as 
\begin{equation}
\label{4.20}m^2=\frac{(D_0-2)(\alpha D_1-2)}{D_1(D-2)}R_1{\left( e^{-2\beta
_c}\right) }^{\frac{D-2}{D_0-2}}\ . 
\end{equation}
For the critical value of $\alpha $ at $\alpha =2/D_1$ the model becomes
degenerated: $U_{eff}\equiv 0$.

As illustration, let $M_1$ be a 3 - dimensional sphere and $a_c=10L_{Pl}$.
This minimum can be achieved for $A={\left( \alpha \pi ^2\right) }^{-1}\cdot
10^{\alpha D_1-2}$. Thus, $\frac 3{2\pi ^2}<A\leq 5\cdot 10^2$ and $%
0<m^2\leq \frac{16}5\cdot 10^{-5}$ for $2/D_1<\alpha \leq 2$ and $D_0=4$. We
see that all conditions (i) - (iii) are satisfied here.

In this section we considered four simple examples of the effective
potential and showed that some of them satisfy conditions (i) - (iii).

%

\section{Internal spaces with two scale factors}

\setcounter{equation}{0}

In this section we extend the consideration of possible excitons from
effective potentials fulfilling conditions (\ref{3.19a}) to internal spaces
with two scale factors. We analyze three potentials --- the pure geometrical
potential, the effective potential of a perfect fluid and the ''monopole''
potential . Stability considerations for Casimir-like potentials can be
found in our paper \cite{22}.

a) pure geometrical potential $U_{eff,0}\equiv U_{eff}(\rho \equiv 0).$

In this case the condition for the existence of an extremum $\frac{\partial
U_{eff,0}}{\partial \beta ^k}=0$ implies a fine-tuning 
\begin{equation}
\label{5.1}\frac{R_k}{D_k}e^{-2\beta _c^k}=\frac{2\Lambda }{D-2}\ ,\quad
k=1,2\qquad \Longrightarrow \qquad e^{\beta _c^k}=\left[ \frac{R_kD_i}{R_iD_k%
}\right] ^{1/2}e^{\beta _c^i}
\end{equation}
of the scale factors and $\mbox{\rm sign}\,\Lambda =\mbox{\rm
sign}\,R_i$. From the Hessian 
\begin{equation}
\label{5.2}
\begin{array}{ll}
a_{(c)ik}\equiv \left. \frac{\partial ^2U_{eff,0}}{\partial \beta
^i\,\partial \beta ^k}\right| _{\vec \beta _c} & =-
\frac{4\Lambda _{eff}}{D_0-2}\left[ \frac{D_iD_k}{D_0-2}+\delta
_{ik}D_k\right]  \\  &  \\  
& =-\frac{4\Lambda }{D-2}\left[ \frac{D_iD_k}{D_0-2}+\delta _{ik}D_k\right]
\exp {\left[ -\frac 2{D_0-2}\sum_{i=1}^2D_i\beta _c^i\right] }
\end{array}
\end{equation}
we see that according to the Sylvester criterion $a_{(c)11}>0\ ,\
a_{(c)22}>0\ ,\ a_{(c)11}a_{(c)22}>a_{(c)12}^2$ there exist massive excitons
for this effective potential in the case of a negative cosmological constant 
$\Lambda <0$ and negative scalar curvatures $R_k<0.$ The masses of the
excitons are easy calculated as eigenvalues of the matrix $\bar G^{-1}A_c$ (%
\ref{3.17}), (\ref{3.18}). Because of 
\begin{equation}
\label{5.3}\bar G^{-1}A_c=\left( 
\begin{array}{cc}
\frac{a_{(c)11}}{D_1}-\frac{a_{(c)11}+a_{(c)12}}{D-2} & \frac{a_{(c)12}}{D_1}%
-\frac{a_{(c)22}+a_{(c)12}}{D-2} \\ \frac{a_{(c)12}}{D_2}-\frac{%
a_{(c)11}+a_{(c)12}}{D-2} & \frac{a_{(c)22}}{D_2}-\frac{a_{(c)22}+a_{(c)12}}{%
D-2}
\end{array}
\right) =-\frac{4\Lambda _{eff}}{D_0-2}\left( 
\begin{array}{cc}
1 & 0 \\ 
0 & 1
\end{array}
\right) 
\end{equation}
they are given as 
\begin{equation}
\label{5.4}
\begin{array}{ll}
m_1^2=m_2^2=-\frac{4\Lambda _{eff}}{D_0-2} & =-
\frac{4\Lambda }{D-2}\exp {\left[ -\frac 2{D_0-2}\sum_{i=1}^2D_i\beta
_c^i\right] } \\  &  \\  
& =2\left| 
\frac{R_1R_2}{D_1D_2}\right| ^{1/2}{\left( e^{-\beta _c^1}\right) }^{\frac{%
2D_1}{D_0-2}+1}{\left( e^{-\beta _c^2}\right) }^{\frac{2D_2}{D_0-2}+1} \\  &
\\  
& =2\left| \frac{R_1}{D_1}\right| \left[ \frac{R_2D_1}{R_1D_2}\right] ^{-
\frac{D_2}{D_0-2}}{\left( e^{-2\beta _c^1}\right) }^{\frac{D-2}{D_0-2}}\ ,
\end{array}
\end{equation}
where the last line follows immediately from the fine-tuning condition (\ref
{5.1}). From eq. (\ref{5.4}) we see that the exciton masses $m_1\ ,\ m_2$ of
the two-scale-factor model are degenerated and related to the corresponding
effective cosmological constant $\Lambda _{eff}$ in the same way as in the
one-scale-factor case (\ref{4.4}). As in the one-scale-factor model, for
specific space configurations the two-scale-factor model allows the
existence of excitons satisfying physical conditions (\ref{3.19a}).

Let us illustrate this situation with an extended version of the example of
section \ref{mark2}a). Suppose that $D_0=4\ ;\ M_1=H^{D_1}/\Gamma _1:\quad
R_1=-D_1(D_1-1)\ ,\ D_1=2\ ,\ a_{(c)1}=10^2L_{Pl}\ \ ;\ M_2=H^{D_2}/\Gamma
_2:\quad R_2=-D_2(D_2-1).$ Mass formula (\ref{5.4}), effective cosmological
constant and fine-tuning condition (\ref{5.1}) read in this case: 
\begin{equation}
\label{5.5} 
\begin{array}{l}
m_1^2=m_2^2=2\cdot (D_2-1)^{-D_2/2}\cdot 10^{-2(D_2+4)}M_{Pl}^2\ , \\ 
\Lambda _{eff}=-(D_2-1)^{-D_2/2}\cdot 10^{-2(D_2+4)}\Lambda _{Pl}\ , \\ 
a_{(c)2}=(D_2-1)^{1/2}a_{(c)1}=(D_2-1)^{1/2}10^2L_{Pl}\ . 
\end{array}
\end{equation}
Thus, conditions (\ref{3.19a}) are fulfilled for internal spaces $M_2$ with
dimensions $D_2\geq D_{2,crit}=40$ . Indeed, in the case of $D_2=40$ we have 
$m_i^2\simeq 2\cdot 10^{-120}M_{Pl}^2\ ,\ \Lambda _{eff}\simeq
-10^{-120}\Lambda _{Pl}\ ,\ a_{(c)2}\simeq 6\cdot 10^2L_{Pl}\ $ and hence
for $D_2>40\ $there hold the relations$\quad m_i\ll M_{Pl}\ ,\left| \Lambda
_{eff}\right| <10^{-120}\Lambda _{Pl}\ ,\quad a_{(c)i}%
\mbox{ \small $^{>}_{\sim}$ }L_{Pl}\ $ as required in (\ref{3.19a}).

b) perfect fluid

For a multicomponent perfect fluid with energy density (\ref{3.30}) the
effective potential reads 
\begin{equation}
\label{5.6}U_{eff}={\left( \prod_{i=1}^2e^{D_i\beta ^i}\right) }^{-\frac
2{D_0-2}}\left[ -\frac 12\sum_{i=1}^2R_ie^{-2\beta ^i}+\Lambda +\kappa
^2\sum_{a=1}^mA_a\exp {\left( -\sum_{k=1}^2\alpha _k^{(a)}D_k\beta ^k\right) 
}\right] \ . 
\end{equation}
Following the same scheme as in the previous considerations we calculate
first extremum condition, Hessian and exciton masses in its general form and
analyze then some concrete subclasses of potentials.

For shortness we introduce the abbreviations 
\begin{equation}
\label{5.7} 
\begin{array}{l}
u_k^{(a)}:=\alpha _k^{(a)}+ 
\frac{2-\sum_{i=1}^2\alpha _i^{(a)}D_i}{D-2}\ ,\quad v_k^{(a)}:=\widetilde{h}%
_a\alpha _k^{(a)}\ ,\quad c_k:=\frac{2\Lambda D_k}{D-2}\ , \\  \\ 
h_a:=\kappa ^2A_ae^{-\alpha _1^{(a)}D_1{\beta _c^1}}e^{-\alpha _2^{(a)}D_2{%
\beta _c^2}}>0\ ,\quad \widetilde{h}_a:=h_a\exp {\left[ -\frac
2{D_0-2}\sum_{i=1}^2D_i\beta _c^i\right] \ .} 
\end{array}
\end{equation}
Extremum condition and Hessian read then 
\begin{equation}
\label{5.8}\frac{\partial U_{eff}}{\partial \beta ^k}=0\ ,k=1,2\quad
\Longrightarrow \quad I_k:=c_k+D_k\kappa
^2\sum_{a=1}^mA_au_k^{(a)}e^{-\alpha _1^{(a)}D_1{\beta _c^1}}e^{-\alpha
_2^{(a)}D_2{\beta _c^2}}-R_ke^{-2\beta _c^k}=0\quad ,\quad k=1,2\ 
\end{equation}
\begin{equation}
\label{5.9}a_{(c)ik}\equiv \left. \frac{\partial ^2U_{eff}}{\partial \beta
^i\,\partial \beta ^k}\right| _{\vec \beta _c}=-\frac{4\Lambda _{eff}}{D_0-2}%
\left[ \frac{D_iD_k}{D_0-2}+\delta _{ik}D_k\right] +\sum_{a=1}^m\widetilde{h}%
_a{\alpha _k^{(a)}D_k\left( {\alpha _i^{(a)}D_i-2\delta _{ik}}\right) \ } 
\end{equation}
and from the auxiliary matrix 
\begin{equation}
\label{5.10}\left[ \bar G^{-1}A_c\right] _{ik}=-\frac{4\Lambda _{eff}}{D_0-2}%
\delta _{ik}+J_{ik}\ ,\quad
J_{ik}=\sum_{a=1}^mv_k^{(a)}(D_ku_i^{(a)}-2\delta _{ik}) 
\end{equation}
we calculate the exciton masses squared as 
\begin{equation}
\label{5.11}m_{1,2}^2=-\frac{4\Lambda _{eff}}{D_0-2}+\frac 12\left[ Tr(J)\pm 
\sqrt{Tr^2(J)-4\det (J)}\right] \ . 
\end{equation}
From eq. (\ref{5.8}) we see that the extremum condition has the form of a
system of equations in variables $z_1=e^{-{\beta _c^1}}\ ,\ z_2=e^{-{\beta
_c^2}}$%
\begin{equation}
\label{5.12}I_k=c_k+D_k\kappa ^2\sum_{a=1}^mA_au_k^{(a)}z_1^{\alpha
_1^{(a)}D_1}z_2^{\alpha _2^{(a)}D_2}-R_kz_k^2=0\quad ,\quad k=1,2 
\end{equation}
and for a given point $p=\left\{ \Lambda ,R_1,R_2,\ A_1,\ldots ,A_m,\alpha
_1^{(1)},\ldots ,\alpha _2^{(m)}\right\} $ in parameter space $\RR
_{par}^{3(m+1)}$ positions of extrema should be found as solutions of this
system. In the general case of $m>1$ and $\alpha _i^{(a)}$ real ($\alpha
_i^{(a)}\in \RR $) this can be done most efficiently by numerical methods.
Partially analytical methods can be applied, e.g. for $\alpha _i^{(a)}$
rational ($\alpha _i^{(a)}\in \QQ $). In this case following representation
holds $\alpha _i^{(a)}D_i=\frac{n_i^{(a)}}{d_i^{(a)}}$ with natural
numerator $n_i^{(a)}\in \NN $ and denominator $d_i^{(a)}\in \NN ^{+}$ ,
where $n_i^{(a)},d_i^{(a)}$ are relative prime, GCD$(n_i^{(a)},d_i^{(a)})=1$%
. Introducing the least common multiple of the denominators $l=$LCM$%
(d_1^{(1)},...,d_2^{(m)})$ and the natural numbers $\vartheta
_i^{(a)}:=\frac l{d_i^{(a)}}n_i^{(a)}$ one has $\alpha _i^{(a)}D_i=\frac{%
\vartheta _i^{(a)}}l$ . Eqs. (\ref{5.12}) transform then to a system of
polynomials 
\begin{equation}
\label{5.13}I_k=c_k+D_k\kappa ^2\sum_{a=1}^mA_au_k^{(a)}y_1^{\vartheta
_1^{(a)}}y_2^{\vartheta _2^{(a)}}-R_ky_k^{2l}=0\quad ,\quad k=1,2 
\end{equation}
in the new variables $y_k=z_k^{1/l}$ , which can be analyzed by algebraic
methods (resultant techniques \cite{34}, algebro-geometrical techniques \cite
{35}) and for rational parameters by methods of number theory \cite{36}. So,
for common roots of equations $I_1=0,\ I_2=0$ the resultants \cite{34} $%
R_{y_1}\left[ I_1,I_2\right] \ ,\ R_{y_2}\left[ I_1,I_2\right] $ must
necessarily vanish 
\begin{equation}
\label{5.14}R_{y_1}\left[ I_1,I_2\right] =w(y_2)=0\ ,\ R_{y_2}\left[
I_1,I_2\right] =w(y_1)=0 
\end{equation}
and the analysis of (\ref{5.12}) can be reduced to an analysis of the
polynomials $w(y_1)\ ,\ w(y_2)$ of degree 
\begin{equation}
\label{5.14a}\deg \left[ w(y_1)\right] \ ,\deg \left[ w(y_2)\right] \leq
\left[ l\stackunder{a}{\max }(\alpha _1^{(a)}D_1+\alpha
_2^{(a)}D_2,2)\right] ^2 
\end{equation}
in only one of the variables $y_1$ and $y_2$ respectively. For explicit
considerations of extremum positions with the help of algebraic methods in
the case of Casimir-like potentials we refer to \cite{22}.

We now turn to the consideration of some concrete subclasses of perfect
fluids.

- b,1) m-component perfect fluid with $\alpha _i^{(a)}=\alpha ^{(a)}$

In this case there exist no massive excitons for vanishing effective
cosmological constants $\Lambda _{eff}=0$. Indeed, $m_{1,2}^2>0$ and eq. (%
\ref{5.11}) imply $Tr(J)>0\ ,\ \det (J)>0$ which with 
\begin{equation}
\label{5.15}J_{ik}=D_kW_1-2\delta _{ik}W_2\ ,\quad
W_1:=\sum_{a=1}^mu^{(a)}v^{(a)}\ ,\ W_2:=\sum_{a=1}^mv^{(a)} 
\end{equation}
read $Tr(J)=D^{^{\prime }}W_1-4W_2>0$ , $\det (J)=2W_2(2W_2-D^{^{\prime
}}W_1)>0$ . But because of $v^{(a)}=\widetilde{h}_a\alpha ^{(a)}>0$ and
hence $W_2>0$ this leads to a contradiction. Thus, for the existence of
massive excitons $m_{1,2}^2>0$ the effective cosmological constant must be
negative $\Lambda _{eff}<0$ .

- b,2) one-component perfect fluid with $\alpha _1\neq \alpha _2$

Again massive excitons are possible for negative effective cosmological
constants $\Lambda _{eff}<0$ only. Here at one hand we have $\det
(J)=-2v_1v_2\delta \frac{D_0-2}{D-2}>0\ ,\ \delta :=D_1\alpha _1+D_2\alpha
_2-2$ and hence $\delta <0$ . On the other hand from $Tr(J)>0$ follows $%
\delta (\alpha _1+\alpha _2-\frac{\delta +2}{D-2})>0$ and hence $%
0>(D_0-2)(\alpha _1+\alpha _2)+D_1\alpha _1+D_2\alpha _2$ . Because of $%
\alpha _k>0$ this is impossible.

- b,3) one-component perfect fluid with $\alpha _1=\alpha _2=\alpha $

For this subclass of b,1) extremum conditions (\ref{5.8}) can be
considerably simplified to yield 
\begin{equation}
\label{5.16}h=\kappa ^2Ae^{-\alpha (D_1{\beta _c^1+}D_2{\beta _c^2)}}=\frac
1{(D_0-2)\alpha +2}\left( \frac{D-2}{D_k}R_ke^{-2\beta _c^k}-2\Lambda
\right) 
\end{equation}
and the same fine-tuning condition as in the case of a pure geometrical
potential 
\begin{equation}
\label{5.17}\tilde C=\frac{R_1}{D_1}e^{-2\beta _c^1}=\frac{R_2}{D_2}%
e^{-2\beta _c^2}\ .
\end{equation}
An explicit estimation of exciton masses and effective cosmological constant
can be easily done. Using (\ref{5.7}), (\ref{5.11}), (\ref{5.15}) we rewrite
the exciton masses squared as 
\begin{equation}
\label{5.18}\left( 
\begin{array}{c}
m_1^2 \\ 
m_2^2
\end{array}
\right) =\frac 1{D-2}\left\{ -4\Lambda +h\left[ (D_0-2)\alpha +2\right]
\left[ \left( 
\begin{array}{c}
D^{^{\prime }}\alpha  \\ 
0
\end{array}
\right) -2\right] \right\} \exp {\left[ -\frac 2{D_0-2}\sum_{i=1}^2D_i\beta
_c^i\right] }\ 
\end{equation}
and transform with (\ref{5.16}) inequalities $m_{1,2}^2>0\ ,\ h>0$ to the
following equivalent condition 
\begin{equation}
\label{5.19}\frac 2{D-2}\Lambda <\tilde C<0\ .
\end{equation}
Hence stable space configurations with massive excitons are only possible
for internal spaces with negative curvature $R_k<0$ . Reparametrizing $%
\Lambda $ according to (\ref{5.19}) as 
\begin{equation}
\label{5.20}\Lambda =\frac{D-2}2\left( \tilde C-\tau \right) \ ,
\end{equation}
with $\tau >0$ --- a new parameter, we get for exciton masses squared and
effective cosmological constant 
\begin{equation}
\label{5.21}\left( 
\begin{array}{c}
m_1^2 \\ 
m_2^2
\end{array}
\right) =\left[ \left( 
\begin{array}{c}
D^{^{\prime }}\alpha \tau  \\ 
0
\end{array}
\right) -2\tilde C\right] \exp {\left[ -\frac 2{D_0-2}\sum_{i=1}^2D_i\beta
_c^i\right] }\ ,
\end{equation}
\begin{equation}
\label{5.22}\Lambda _{eff}=-\frac{D_0-2}2\left[ \tau \frac{(D-2)\alpha }{%
(D_0-2)\alpha +2}-\tilde C\right] \exp {\left[ -\frac
2{D_0-2}\sum_{i=1}^2D_i\beta _c^i\right] \ .}
\end{equation}
According to definition (\ref{5.20}) and equations (\ref{5.16}), (\ref{5.17}%
) the parameter $\tau $ can be expressed in terms of $\tilde C$ and $R_k$ as 
\begin{equation}
\label{5.22a}\tau =\kappa ^2A\frac{(D_0-2)\alpha +2}{D-2}\left| \tilde
C\right| ^{\frac{D^{^{\prime }}\alpha }2}\prod_{k=1}^2\left| \frac{D_k}{R_k}%
\right| ^{\frac{D_k\alpha }2}\ .
\end{equation}
Comparison of equations (\ref{5.21}), (\ref{5.22}) with formula (\ref{5.4})
shows that for $\tau \ll \tau _0\equiv \left| \tilde C\right| \min (\frac
2{D^{^{\prime }}\alpha },\frac{(D_0-2)\alpha +2}{(D-2)\alpha })$ we return
to the pure geometrical potential considered in paragraph a). So physical
conditions (\ref{3.19a}) are fulfilled for internal space configurations
with sufficiently high dimensions greater then some critical dimension $%
D_{crit}$. From (\ref{5.21}) and (\ref{5.22}) we see that depending on the
value of $\tau $ this critical dimension $D_{crit}$ can only be larger then
that for the pure geometrical model. According to (\ref{5.22a}) there exist
excitons for any positive and finite values of the fluid parameter $A$ , but
than larger $A$ for fixed $\alpha $ than larger would be the critical
dimension $D_{crit}$ . (Here we take into account that $\kappa ^2=\mu $ and
that the volume $\mu $ of the compact internal factor spaces with constant
negative curvature is finite.)

Comparing the results of this subsection with the results of section \ref
{mark2}d) we see that there exists a different behavior of the perfect fluid
models in the case of vanishing effective cosmological constant $\Lambda
_{eff}=0$ . For the one-scale-factor model massive excitons are allowed for $%
\Lambda _{eff}=0$, whereas in the two-scale-factor model they cannot occur.
An explanation of this situation will be given in the next section.

- c) ''monopole'' potential: $\rho =\sum_{k=1}^2(f_k)^2e^{-2D_k\beta ^k}$

For the ''monopole'' potential the extremum condition (\ref{3.1}) leads in
the case of vanishing effective cosmological constant $\Lambda _{eff}=0$ to
a fine-tuning of the scale factors 
\begin{equation}
\label{5.23}\frac{R_k}{2D_k(f_k)^2}=e^{-2\beta ^k(D_k-1)}
\end{equation}
and 
\begin{equation}
\label{5.24}\Lambda =\frac 12\sum_{k=1}^2R_ke^{-2\beta ^k}\frac{D_k-1}{D_k}\
,
\end{equation}
so that, as for the one-scale-factor model, extrema are only possible iff $%
R_k>0\ ,\ \Lambda >0$ . Because the ''monopole'' potential formally
coincides with the potential of a perfect fluid with parameters $\alpha
_k^{(a)}=2\delta _{ak}$ the exciton masses are given by eq. (\ref{5.11}) 
\begin{equation}
\label{5.25}m_{1,2}^2=\frac 12\left[ Tr(J)\pm \sqrt{Tr^2(J)-4\det (J)}%
\right] \ ,
\end{equation}
where in terms of abbreviations (\ref{5.7}) matrix $J$ reads 
\begin{equation}
\label{5.26}J_{ik}=4\widetilde{h}_k(D_k-1)\left[ \delta _{ik}-\frac{D_k}{D-2}%
\right] \ .
\end{equation}
One immediately verifies that $Tr(J)>0\ ,\ \det (J)>0\ ,\ Tr^2(J)-4\det
(J)\geq 0$ for dimensions $D_1>1\ ,\ D_2>1$ and hence $0<m_2^2\leq \frac
12Tr(J)\leq m_1^2<Tr(J)$ . This means that physical conditions (\ref{3.19a})
are satisfied if $Tr(J)\leq M_{Pl}^2$ and $e^{\beta _c^k}$%
\mbox{ \small
$^{>}_{\sim}$ }\thinspace $L_{Pl}$\ . Substituting 
\begin{equation}
\label{5.27}\widetilde{h}_k=\frac{R_k}{2D_k}e^{-2\beta _c^k}\exp {\left[
-\frac 2{D_0-2}\sum_{i=1}^2D_i\beta _c^i\right] }
\end{equation}
into (\ref{5.26}) we get the matrix trace as 
\begin{equation}
\label{5.28}Tr(J)=\frac 2{D-2}\left[ {\sum_{k=1}^2\frac{(D_k-1)}{D_k}%
R_k(D-2-D_k)}e^{-2\beta _c^k}\right] \exp {\left[ -\frac
2{D_0-2}\sum_{i=1}^2D_i\beta _c^i\right] \ .}
\end{equation}
With this formula at hand we have e.g. for an internal space configuration $%
M_1\times M_2:\ M_1=$ $S^3\ ,\ a_{(c)1}=10L_{Pl}\ ;\ \ M_2=$ $S^5\ ,\
a_{(c)2}=10^2L_{Pl}$ the\ estimate $Tr(J)\approx 56\cdot 10^{-14}M_{Pl}^2\ll
M_{Pl}^2$ and all conditions (i) - (iii) of (\ref{3.19a}) are satisfied.


\section{Exciton masses and scale factor constraints}

\setcounter{equation}{0}

In this section we derive a relation between the exciton masses $m_{(c)1},\
m_{(c)2}$ of a model with two independently varying scale factors $\beta
^1,\beta ^2$ and the effective mass $m_{(c)0}$ of the exciton which occurs
under scale factor reduction, i.e. when the scale factors of the model are
connected by a constraint $\beta =\beta ^1=\beta ^2$ .

In order to simplify our calculation we introduce the projection operator $P$
on the constraint subspace $\RR _P^1=\left\{ \bar \beta =(\beta ^1,\beta
^2)\mid \beta ^1-\beta ^2=\bar a\cdot \bar \beta =0\ ,\ \bar
a=(1,-1)\right\} $ of the 2-dimensional target space $\RR _T^2$ of the $%
\sigma -$model 
\begin{equation}
\label{6.1}P\RR _T^2=\RR _P^1\subset \RR _T^2\ . 
\end{equation}
Explicitly this projection operator can be constructed from the normalized
base vector $\bar e$ of the subspace \mbox{$\RR _P^1\ .$} With $\bar e=\frac
1{\sqrt{2}}\left( 
\begin{array}{c}
1 \\ 
1 
\end{array}
\right) $ we have 
\begin{equation}
\label{6.2}P=\bar e\otimes \bar e^{^{\prime }}=\frac 12\left( 
\begin{array}{c}
1 \\ 
1 
\end{array}
\right) \otimes \left( 
\begin{array}{cc}
1 & 1 
\end{array}
\right) =\frac 12\left( 
\begin{array}{cc}
1 & 1 \\ 
1 & 1 
\end{array}
\right) 
\end{equation}
and $P^2=P\ ,\ P\bar a=0$ .

Let us now calculate the exciton mass $m_{(c)0}$ for the reduced model. For
this purpose we introduce the exciton Lagrangian, written according to
section \ref{mark1} in terms of the fluctuation fields $\bar \eta =(\eta
^1,\eta ^2),\ \eta ^i\equiv \beta ^i-\beta _c^i$ 
\begin{equation}
\label{6.3}{\cal L}_{exci}=-\left[ \bar \eta \bar G\widehat{K}\bar \eta
+\bar \eta A_{(c)}\bar \eta \right] \ .
\end{equation}
$\widehat{K}:=\overleftarrow{\partial }_\mu \widehat{g}^{(o)\mu \nu }
\overrightarrow{\partial }_\nu $ denotes the pure kinetic operator. Under
scale factor reduction $\bar \eta =(\eta ,\eta )$ this Lagrangian transforms
to 
\begin{equation}
\label{6.4}{\cal L}_{exci}=-\left[ \gamma _1\eta \widehat{K}\eta +\gamma
_{(c)2}\eta ^2\right] \ ,\quad 
\end{equation}
\begin{equation}
\label{6.5}\gamma _1:=2\bar e^{^{\prime }}\bar G\bar e=\sum_{i,j}\bar
G_{ij}\ ,\quad \gamma _{(c)2}:=2\bar e^{^{\prime }}A_{(c)}\bar
e=\sum_{i,j}A_{(c)ij}
\end{equation}
so that the substitution $\eta =\gamma _1^{-1/2}\psi $ yields the effective
one-scale-factor Langrangian \\ ${\cal L}_{exci}=-\left[ \psi \widehat{K}%
\psi +\psi m_{(c)0}^2\psi \right] $ with exciton mass $m_{(c)0}^2=\gamma
_{(c)2}/\gamma _1$ . Taking into account that $\bar e^{^{\prime
}}A_{(c)}\bar e=Tr\left[ PA_{(c)}\right] $ , \ $A_{(c)}=Q^{^{\prime
}}S_c^{^{\prime }}M_{(c)}^2S_cQ$ and $M_{(c)}^2=diag(m_{(c)1}^2,m_{(c)2}^2)$
the needed relation between the exciton masses of the reduced and unreduced
two-scale-factor models is now easily established as 
\begin{equation}
\label{6.6}m_{(c)0}^2=2\gamma _1^{-1}Tr\left[ QPQ^{^{\prime }}S_c^{^{\prime
}}M_{(c)}^2S_c\right] .
\end{equation}
With the use of 
\begin{equation}
\label{6.7}QPQ^{^{\prime }}=\frac 12D^{^{\prime }}\frac{D-2}{D_0-2}\left( 
\begin{array}{cc}
1 & 0 \\ 
0 & 0
\end{array}
\right) \ ,\quad \gamma _1=D^{^{\prime }}\frac{D-2}{D_0-2}
\end{equation}
and the $SO(2)-$rotation matrix $S_c$ from (\ref{3.15}), (\ref{3.16}) this
formula can be considerably simplified to give the final relation 
\begin{equation}
\label{6.8}m_{(c)0}^2=\cos {}^2(\alpha _c)m_{(c)1}^2+\sin {}^2(\alpha
_c)m_{(c)2}^2\ .
\end{equation}

In its compact form this mass formula implicitly reflects the behavior of
the effective potential $U_{eff}$ in the vicinity $\Omega _{\vec \beta
_c}\subset \RR _T^2$ of the extremum point $\vec \beta _c$. So, the exciton
masses squared $m_{(c)1}^2,\ m_{(c)2}^2$ describe the potential as function
over the two-dimensional $\vec \beta _c-$vicinity $\Omega _{\vec \beta _c}$
, whereas $m_{(c)0}^2$ characterizes $U_{eff}$ as function over the line
interval $\Omega _{\vec \beta _c}\cap \RR _P^1$ only. Comparison of the
minimum conditions of the unreduced and reduced two-scale-factor model 
\begin{equation}
\label{6.9}
\begin{array}{ll}
m_{(c)1,2}^2>0\ :\  & a_{(c)11}>0\ ,\ a_{(c)22}>0\ , \\  
& a_{(c)11}\cdot a_{(c)22}>\left( a_{(c)12}\right) ^2
\end{array}
\end{equation}
and 
\begin{equation}
\label{6.10}m_{(c)0}^2>0\ :\quad (a_{(c)11}+a_{(c)22}+2a_{(c)12})>0
\end{equation}
shows that stable configurations of reduced models with $m_{(c)0}^2>0$ are
not only possible for stable configurations of the unreduced model $%
m_{(c)1}^2>0\ ,\ m_{(c)2}^2>0\ $ , but even in cases when the potential $%
U_{eff}$ has a saddle point at $\vec \beta _c$ and the unreduced model is
unstable. For the masses we have in these cases $m_{(c)1}^2>0\ ,\
m_{(c)2}^2<0\ $ or $m_{(c)1}^2<0\ ,\ m_{(c)2}^2>0\ $ and massive excitons in
the reduced model correspond to exciton - tachyon configurations in the
unreduced model.

\section{Conclusions}

\setcounter{equation}{0}

This paper was devoted to the problem of stable compactification of internal
spaces. This is one of the most important problems in multidimensional
cosmology, because via stable compactification of the internal dimensions
near Planck length we can explain unobservability of extra-dimensions. With
the help of dimensional reduction we obtained an effective four -
dimensional theory in Brans - Dicke and Einstein frames. The Einstein frame
was considered here as a physical one \cite{37}. In this frame we derived an
effective potential. It was shown that small excitations of the scale
factors of internal spaces near minima of the effective potential have a
form of massive scalar particles (gravitational excitons) developing in the
external space - time. Detection of these excitations can prove the
existence of extra - dimensions. Particular examples of effective potentials
were investigated in the one - and two - internal - space cases. Parameters
of the models which ensure a minimum were obtained and masses of the
excitons were estimated. The solutions at the minima of the potential are
stable against small perturbations of the scale factor(s) of the expanding
external universe \cite{12}. We would like to note, that the problem of
stable compactification in MCM with more than one internal scale factor was
considered first for pure geometrical models in papers \cite{14,15}.
However, the analysis of the effective potential minima existence was not
complete there.

Our analysis shows that conditions for the existence of stable
configurations may be quite different for one- and two-scale-factor models.
For example, in the case of a one-scale-factor model which is filled with a
one-component perfect fluid stable compactifications are possible for
vanishing effective cosmological constant $\Lambda _{eff}=0$ and parameters $%
\alpha \ $ from the restricted interval$\ 2/D_1<\alpha \le 2$ determining
the equation of state in the internal space: $P_1=(\alpha -1)\rho $ . In the
case of two-scale-factor models stable compactifications can exist for
negative effective cosmological constants $\Lambda _{eff}<0$ only, but for
values of the parameter $\alpha $ from the usual interval : $0\le \alpha \le
2$ (here, $\alpha $ determines the equations of state in both internal
spaces : $P_1=(\alpha -1)\rho ,P_2=(\alpha -1)\rho $ ). At first sight the
difference in the behavior of these two models looks a bit strange because
the one-scale-factor model can be obtained by reduction of the
two-scale-factor model with the help of the constraint $\beta _1=\beta
_2\equiv \beta $. As it was shown in section 6, such a different behavior
may take place because stable configurations of reduced models are not only
possible for stable configurations of unreduced models, but even in cases
when the effective potential $U_{eff}$ of the unreduced model has a saddle
point. In the case of our two-scale-factor model with one-component perfect
fluid we get such a saddle point for configurations with $\Lambda _{eff}=0$
and $2/(D_1+D_2)<\alpha \le 2$ .

In the present paper we did not consider the case of degenerated minima of
the effective potential, for example, self - interaction - type potentials
or Mexican - hat - type potentials. In the former case one obtains massless
fields with self - interaction. In the latter case one gets massive fields
together with massless ones. Here, massless particles can be understood as
analog of Goldstone bosons. This type of the potential was described in \cite
{19}.

Another possible generalization of our model consist in the proposal that
the additional potential $\rho $ may depend also on the scale factor of the
external space. It would allow, for example, to consider a perfect fluid
with arbitrary equation of state in the external space. \bigskip


{\bf Acknowledgements}\\ We thank V.Melnikov, K.Bronnikov, V.Ivashchuk and
V.Gavrilov for useful discussions during the preparation of this paper and
the referee for drawing our attention to letter \cite{24a}. UG acknowledges
financial support from DAAD (Germany).

\end{document}